\documentclass[aps,prb,showpacs,groupedaddress,twocolumn]{revtex4}

\usepackage{graphicx}
\begin{document}

\title{disorder-induced enhancement of transport through graphene p-n
junctions}

\author{Wen Long$^{1}$, Qing-feng Sun$^{2,\star}$, and Jian
Wang$^{3}$} \affiliation{ $^1$Department of Physics, Capital
Normal University, Beijing
100037, China\\
$^2$Institute of Physics, Chinese
Academy of Sciences, Beijing 100190, China\\
$^3$Department of Physics and the center of theoretical and
computational physics, The University of Hong Kong, Hong Kong,
China
 }

\date{\today}

\begin{abstract}
We investigate the electron transport through a graphene p-n
junction under a perpendicular magnetic field. By using
Landauar-Buttiker formalism combining with the non-equilibrium Green
function method, the conductance is studied for the clean and
disordered samples. For the clean p-n junction, the conductance is
quite small. In the presence of disorders, it is strongly enhanced
and exhibits plateau structure at suitable range of disorders. Our
numerical results show that the lowest plateau can survive for a
very broad range of disorder strength, but the existence of high
plateaus depends on system parameters and sometimes can not be
formed at all. When the disorder is slightly outside of this
disorder range, some conductance plateaus can still emerge with its
value lower than the ideal value. These results are in excellent
agreement with the recent experiment.\cite{ref7}
\end{abstract}

\pacs{73.63.-b, 73.23.-b, 81.05.Uw, 73.21.Hb}

\maketitle

Due to the recent success in fabrication of graphene, a single-layer
hexagonal lattice of carbon atoms, great attention has been
attracted in the research of graphene.\cite{ref1,ref2} The unique
band structure of graphene with a linear dispersion relation ($E=\pm
\hbar v|{\bf k}|$) near the Dirac-points leads to many peculiar
properties.\cite{ref3} For instance, the quasi-particles obey the
Dirac-like equation and have the relativistic-like behaviors with
zero rest mass. Its Hall plateaus assume the half-integer values $g
(n+1/2) e^2/h$ with the degeneracy $g=4$.\cite{ref1,ref2} By varying
the gate voltage, the carrier of graphene can be tuned from
electron-like to hole-like and vice versa. To examine the interplay
between the electron-like and hole-like quasi-particles, a graphene
p-n junction would be a good candidate. Many exciting phenomena
reflecting the massless Dirac character of
carriers,\cite{ref4,ref5,ref6} such as relativistic Klein
tunneling\cite{ref4} and Veselago lensing\cite{ref5}, were predicted
for the graphene p-n junction.

Very recently, the graphene junction has been realized
experimentally.\cite{ref7} As expected, it was found that in quantum
Hall regime the two-terminal conductance exhibits quantized plateaus
with half-integer values for the p-p or n-n junctions. For the
disordered p-n junction, new plateaus emerge at $e^2/h$ and $(3/2)
e^2/h$. At about the same time, a theoretical analysis qualitatively
explained the appearance of these plateaus that is due to the
mixture of the electron and hole Hall edge modes in the p-n
boundary.\cite{ref8} After that, subsequent works have also
investigated the graphene p-n junction.\cite{ref9} However, these
theory can not accounted for the experimentally observed plateau
that appeared at about $1.4 e^2/h$ which is lower than the expected
value $(3/2) e^2/h$. In addition, the reason that the expected
plateaus at $3 e^2/h$ or higher values have not been observed
remained mysterious. In view of this situation, it is clear that a
quantitatively analysis for the graphene p-n junction is needed.

In this paper, we theoretically study the electron transport through
the p-n junction of disordered graphene under a perpendicular
magnetic field $B$. By using the tight-binding model and the
Landauar-Buttiker formalism combining with the non-equilibrium Green
function method, the conductance is calculated in both clean and
disordered samples. Numerical results show that the conductance is
very weak in the clean p-n region at large $B$. With increasing of
the disorder, the conductance is strongly enhanced and new plateaus
emerge at $e^2/h$, $(3/2) e^2/h$, etc. The range of the disorder
strength $W$ needed for the existence of the lowest plateau $e^2/h$
is very broad, so this plateau can easily be observed. But for the
plateaus corresponds to higher quantization values, the range of $W$
can be very narrow. Sometimes these higher order plateaus can not be
formed. Our results seem to suggest that the higher the plateau
value, the more difficult it is to observe experimentally. When the
disorder is slightly off this disorder range, the conductance
plateaus can still emerge, but its value is lower than the expected
one. These results are in excellent agreement with experimental
data.

In the tight-binding representation, the Hamiltonian of the graphene
p-n junction (see Fig.1a) is given by:\cite{ref10}
\begin{equation}
  H = \sum\limits_i \epsilon_i a^{\dagger}_i a_i
      -\sum\limits_{<ij>} t e^{i\phi_{ij}} a_i^{\dagger} a_j
\end{equation}
where $a_i^{\dagger}$ and $a_i$ are the the creation and
annihilation operators at the discrete site $i$, and $\epsilon_i$ is
the on-site energy. In the left and right leads, $\epsilon_i=E_L$ or
$E_R$, which can be controlled by the gate voltages. The disorder
exists only in the center region. The potential drop from the right
to the left leads is assumed to be linear, i.e., $\epsilon_i=
k(E_R-E_L)/(2M+2)+E_L +w_i$, where $M$ is the length of the center
region and $k=0,1,2,...2M+1$ (see Fig.1a). The on-site disorder
energy $w_i$ is uniform distributed in the range $[-W/2, W/2]$ with
the disorder strength $W$. The size of the center region is
described by the width $N$ and length $M$, and it has $2N(2M+1)$
carbon atoms. In Fig.1a, it shows a system with $N=2$ and $M=4$.
Here we only consider the zigzag edge graphene, but all results are
similar for the armchair edge graphene. The second term in
Hamiltonian (1) describes the nearest neighbor hopping. Due to the
existence of the perpendicular magnetic field $B$, a phase
$\phi_{ij}$ is added in the hopping element, and $\phi_{ij}=\int_i^j
\vec{A} \cdot d\vec{l}/\phi_0$ with the vector potential
$\vec{A}=(-By,0,0)$ and $\phi_0=\hbar/e$.

The current flowing through the graphene p-n junction is calculated
from the Landauer-B$\ddot{u}$ttiker formula:\cite{ref11} $I=(2e/h)
\int d \epsilon ~ T_{LR}(\epsilon)[f_L(\epsilon)-f_R(\epsilon)]$,
where $f_{\alpha}(\epsilon) =1/\exp[(\epsilon-eV_{\alpha})/k_BT+1]$
($\alpha=L,R$) is the Fermi distribution function in the left and
right graphene lead. Here $T_{LR}(\epsilon) = Tr [{\bf \Gamma}_L
{\bf G}^r {\bf \Gamma}_R {\bf G}^a ]$ is the transmission
coefficient, where the linewidth functions ${\bf
\Gamma}_{\alpha}(\epsilon)= i[{\bf\Sigma}^r_{\alpha}(\epsilon)
-{\bf\Sigma}^a_{\alpha}(\epsilon)]$, the Green functions ${\bf
G}^r(\epsilon)=[{\bf G}^{a}(\epsilon)]^{\dagger} =1/[\epsilon-{\bf
H}_{cen}-{\bf\Sigma}^r_L -{\bf\Sigma}^r_R]$, ${\bf H}_{cen}$ is the
Hamiltonian in the center region, and
${\bf\Sigma}^r_{\alpha}(\epsilon)$ is the retarded self-energy due
to the coupling to the lead-$\alpha$ that can be calculated
numerically\cite{ref12}. After obtaining the current $I$, the linear
conductance is given by $G=\lim\limits_{V \rightarrow 0} dI/dV$.

In the following numerical calculations, we use the hopping energy
$t\approx 2.75eV$ as the energy unit. Since the hoping energy $t$
corresponds to $10^4 K$, we can safely set the temperature to zero
in our calculation. The width $N$ is chosen as $N=50$ in all
calculations. Since the nearest-neighbor carbon-carbon distance
$a=0.142nm$, the width is $3aN=21.6nm$ for $N=50$. The magnetic
field is expressed in terms of $\phi$ with $\phi \equiv
(3\sqrt{3}/4) a^2 B/\phi_0$ and $(3\sqrt{3}/2) a^2 B$ is the
magnetic flux in the honeycomb lattice. In the presence of the
disorder, the conductance is averaged over up to 2000 random
configurations except for Fig.2b where only 400 random
configurations were used for each data. In the experiment, the
typical concentration of electrons or holes is around $10^{13}/cm^2$
that corresponds to the on-site energies $E_L,E_R \le 0.1t$. So we
will mainly focus on the region of $E_L$ and $E_R$ within $0.3t$. In
this range of energy, the dispersion relation is linear and exhibits
Dirac behaviors.

We first study the clean graphene junction. Fig.1b and c show the
conductance $G$ versus the Fermi level of right lead $E_R$ setting
the magnetic field $B=0$. In the n-n region with $E_L,E_R<0$, $G$ is
approximatively quantized and exhibits a series of equidistant
plateaus at the half-integers (in the unit of $4e^2/h$) due to the
transverse sub-bands of the lead with finite width. Because of the
linear dispersion relation the transverse sub-bands $E_n$ of the
confined graphene are in equidistant instead of $E_n \sim n^2$ of
the usual two-dimensional electron gas. While for $E_R<E_L$, due to
the fixed sub-band numbers in the left region, no more higher
plateaus appear. On the other hand, in the p-n region with $E_L<0$
and $E_R>0$, there is no plateaus. The conductance $G$ in the p-n
region ($E_R>0$) is always less than the corresponding plateau value
in the n-n region ($E_R<0$). Due to the occurrence of the Klein
tunneling processes\cite{ref4} the conductance is quite large, e.g.,
$G>e^2/h$ for almost all positive $E_R$ at $M=5$. With the increase
of $M$, the Klein tunneling processes are slightly weakened and so
is the conductance.

Next, we examine the effect of the magnetic field $B$ in the clean
sample. With the increase of $B$, the equidistant sub-bands
gradually evolve into the Landau levels which scales as $E_n \propto
\sqrt{n}$ for the Dirac particle. The conductance plateaus in the
n-n region evolve into the Hall plateau, and the conductance $G$ in
the p-n region is strongly suppressed at small $E_R$. Fig.1d,e and
Fig.2a show $G$ at a high magnetic field with the magnetic flux (or
phase) $\phi=0.007$. We see perfect Hall plateau in the n-n junction
with equidistant in the scale of $E_R^2$ (see Fig.2a). The plateau
values are given by $\min(|\nu_L|,|\nu_R|)2e^2/h$ where
$\nu_{\alpha}$ is the filling factors in the lead-$\alpha$. In
particular the Hall plateaus, i.e. the curve of $G$-$E_R$ in Fig.1d
and 1e at $E_R<0$, do not depend on $M$. However, in the p-n region
($E_R>0$ and $E_L<0$), no plateaus exist. The conductance $G$ is
small and strongly depends on the junction length $M$. For the small
filling factors $\nu_L$ and $\nu_R$ or the large junction length
$M$, $G$ is almost zero. This is because for the clean p-n junction,
the Hall edge states for electrons and holes are well separated in
the space and do not form the mixture of the states leading to the
very small conductance.

In the following, we shall focus on how the conductance $G$ is
affected by the disorders. Fig.2 plots $G$ versus the energy $E_L$
and $E_R$ with the disorder strength $W=0$ and $W=2$. In the
presence of disorders, the conductance $G$ in the p-n region (or n-p
region) is strongly enhanced due to the mixture of the electron and
hole Hall edge states, while $G$ in the n-n and p-p regions are
slightly weakened. At fixed filling factors $\nu_L$ and $\nu_R$, $G$
is approximatively a constant. As $\nu_L$ or $\nu_R$ varies, a jump
occurs in $G$ with the borders between $\nu_L$ and $\nu_R$ regions
clearly seen in Fig.2b.

Now we investigate the effect of disorders on the conductance in
more detail. Fig.3 depicts the conductance vs $E_R$ at fixed
$E_L=-0.1$ ($\nu_L=-2$) and $-0.2$ ($\nu_L=-6$). When $W=0$, $G$ is
small in the p-n region and $G$ exhibits the Hall plateaus in the
n-n region. With the increase of $W$ from $0$, the conductance $G$
in the p-n region is strongly enhanced even for very small $W$. For
example, for $W=0.02$ or $W=0.05$, $G$ is greater than $0.2 e^2/h$,
which is much larger than that ($G<0.001 e^2/h$) at $W=0$ (see
Fig.3a,c). When $W=0.1$, the lowest conductance plateau with
$\nu_L=-2$ and $\nu_R=2$ is well established with its plateau value
at $e^2/h$. In particular, this plateau remains for a broad range of
disorder strength $W$ (from 0.1 to 3). For the higher filling
factors, the conductance is also enhanced by the disorder, but it
requires much large disorder to reach its ideal plateau value at
$[|\nu_L||\nu_R|/(|\nu_L|+|\nu_R|)] e^2/h$. For example, for
$\nu_L=-2$ and $\nu_R=6$ or $\nu_L=-6$ and $\nu_R=2$, the
conductance reaches the plateau of $(3/2) e^2/h$ when $W=2$ (see
Fig.3b,d). In the n-n region, the Hall plateau is not affected by
the small disorders and kept their values at $\min(|\mu_L|,|\mu_R|)
e^2/h$. If the disorder strength $W$ is increased further, the
conductance $G$ starts to drop in both n-n and p-n regions. For very
large $W$ (e.g. $W=6$ or larger), the system enters the insulating
regime and $G$ is very small for all $E_L$ and $E_R$. Here we wish
to emphasize two points: (i). We have seen that the new plateau
survives only within certain range of $W=[W_{min},W_{max}]$. When
the disorder is slightly below $W_{min}$ or above $W_{max}$, the
conductance $G$ still exhibits a plateau but its value is less than
the value of ideal plateau. For example, the plateau of $\nu_L=-2$
and $\nu_R=6$ is less than $(3/2)e^2/h$ when $W=1$ and $W=3$ (see
Fig.3c and 3d).
(ii). For some high filling factor region (e.g. $\nu_L=-6$ and
$\nu_R=6$), the conductance plateau does not emerge at all for any
$W$. Because it is much more difficult to completely mix all states
for high filling factor case, so the system goes to the insulating
regime before the occurrence of the complete state mixture. These
numerical results are in excellent agreement with the
experiment.\cite{ref7}

We now focus on the conductance vs disorder strength for energies
$E_L$ and $E_R$ shown in Fig.2a (solid dots). In the n-n region,
the Hall edge states are very robust against disorders so the
conductance remains quantized at small $W$ (see Fig.4b). At large
disorders, the edge states are destroyed and the Hall conductance
monotonically decreases with increasing of $W$. In the p-n region
(Fig.4a,c and d), the enhancement of conductance due to the states
mixing at moderate disorders is clearly seen. For the lowest
filling factors with $\nu_L=-2$ ($E_L=-0.1$) and $\nu_R=2$
($E_R=0.1$), $G$ reached its ideal plateau value $e^2/h$ at
$W=0.09$ and stayed there until $W=3$ (see Fig.4a).
We emphasize that this range of disorder $W$ (from 0.09 to 3) is
very broad, extends in almost two orders of magnitude! So the lowest
plateau can easily be observed experimentally. For $\nu_L=-2$ and
$\nu_R=6$ ($E_R=0.2$) (or $\nu_R=10$ with $E_R=0.25$), the left side
of the sample has an electron Hall edge state and the right side has
three (or five) holes Hall edge states. As a result of the mixture
of the left-side electron state and one of right-side hole states,
the conductance $G$ develops a step around $e^2/h$. Upon further
increasing $W$, the complete mixture of the left-side electron state
and all right-side hole states occurs at $W=1.6$ (or $1.7$) and the
conductance $G$ reaches the ideal plateau value $(3/2)e^2/h$ (or
$(5/3) e^2/h$). Note that this ideal plateau exists only within the
disorder window $1.7<W<2.6$ (or $1.6<W<2.4$) that is much narrower
than that of the lowest plateau. Our results also show that for the
case of higher filling factors (e.g. $\nu_L=-6$ and $\nu_R=6$), the
ideal plateau $3e^2/h$ can not be reached for any disorders. When
the center region becomes longer or shorter, the conductance $G$
shows similar results (see Fig.4c,d). For a short center region
(e.g. $M=10$), the conductance reaches the ideal lowest plateau at a
larger $W$ with a smaller plateau width. The high conductance
plateau at $\nu_L=-6$ and $\nu_R=6$ also appears. On the other hand,
for a longer center region (e.g. $M=40$), $G$ reaches the ideal
lowest plateau at a smaller $W$ with a wider plateau. Except for the
lowest plateau, all other plateaus (including $(3/2)e^2/h$ and
$(5/3)e^2/h$) do not appear for $M=40$.

Finally, we study the conductance fluctuation $rms(G)\equiv
\sqrt{\langle G-\langle G\rangle\rangle^2}$, where
$\langle...\rangle$ is the average over the disorder configurations
with the same disorder strength $W$. Fig.5 shows $rms(G)$ versus $W$
with the same set of parameters as in Fig.4a and 4b. In the n-n
region (see Fig.5b), there is no fluctuation of the Hall edge states
at small $W$. When disorder increases, the conductance fluctuates
when the edge states are partially destroyed. At large disorders,
$rms(G)$ eventually goes to zero and enters the insulating regime.
On the other hand, in the p-n region (see Fig.5a), the fluctuation
$rms(G)$ is small for both small and large $W$. But $rms(G)$ is
large for intermediate $W$ and usually exhibits a double-peak
structure. In particular, $rms(G)$ does not have the plateau
although the conductance has a very long plateau especially at
$\nu_L=-2$ and $\nu_R=2$.

In summary, the electron transport through a graphene p-n junction
under the perpendicular magnetic field is numerically and
quantitatively studied. We find the conductance is quite small for
the clean p-n junction. But the disorder can drastically enhance the
conductance leading to the conductance plateaus. The lowest
conductance plateaus can sustain for a very broad range of disorder
strength (about two orders of magnitude), but the higher plateaus
are difficult to form. When the disorder is slightly outside of this
disorder range, some conductance plateaus in $G$ vs $E_R$ curve can
also emerge with plateau value smaller than the ideal value.

{\bf Acknowledgments:} We gratefully acknowledge the financial
support from a RGC grant from the Government of HKSAR grant number
HKU 7044/05P (J.W.); NSF-China under Grant Nos. 10525418 and
10734110 (Q.F.S.).

\newpage

\begin{figure}
\caption{ (Color online) (a). The schematic diagram for a zigzag
edge graphene p-n junction. (b), (c), (d), and (e): the conductance
$G$ vs $E_R$ for different center lengths $M$ at $W=0$. The
parameters $E_L=-0.1$ for (b) and (d) and $E_L=-0.2$ for (c) and
(e), and $\phi=0$ for (b) and (c) and $\phi=0.007$ for (d) and (e).
 } \label{fig:1}
\end{figure}

\begin{figure}
\caption{ (Color online) The conductance $G$ (in the unit of
$2e^2/h$) vs $E_L$ and $E_R$ with $M=20$, $\phi=0.007$. (a): $W=0$
and (b): $W=2$.
 }
\label{fig:2}
\end{figure}

\begin{figure}
\caption{ The conductance $G$ vs $E_R$ for the different disorder
strengths $W$, with the parameters $M=20$, $\phi=0.007$. (a) and
(b): $E_L=-0.1$. (c) and (d): $E_L=-0.2$.
 } \label{fig:3}
\end{figure}

\begin{figure}
\caption{ The conductance $G$ vs disorder strength for $E_L$ and
$E_R$ fixed at different points shown in Fig.2a with $\phi=0.007$.
The system size, $M=20$ for (a) and (b), $M=10$ for (c), and $M=40$
for (d). $E_L$ and $E_R$ in (a) and (c) are same as in (d). }
\label{fig:4}
\end{figure}

\begin{figure}
\caption{ $rms(G)$ vs $E_R$. The parameters in (a) and (b) are the
same as in Fig.4a and 4b, respectively.
 } \label{fig:5}
\end{figure}

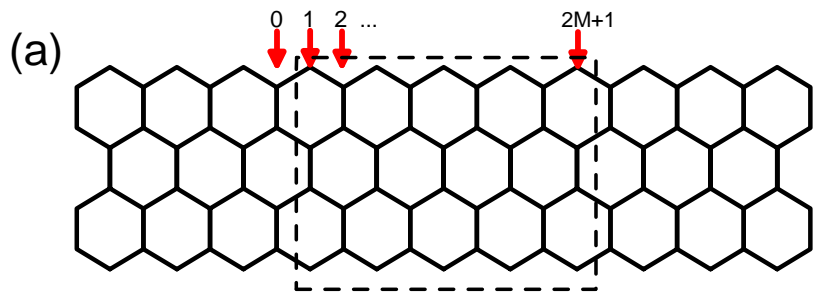
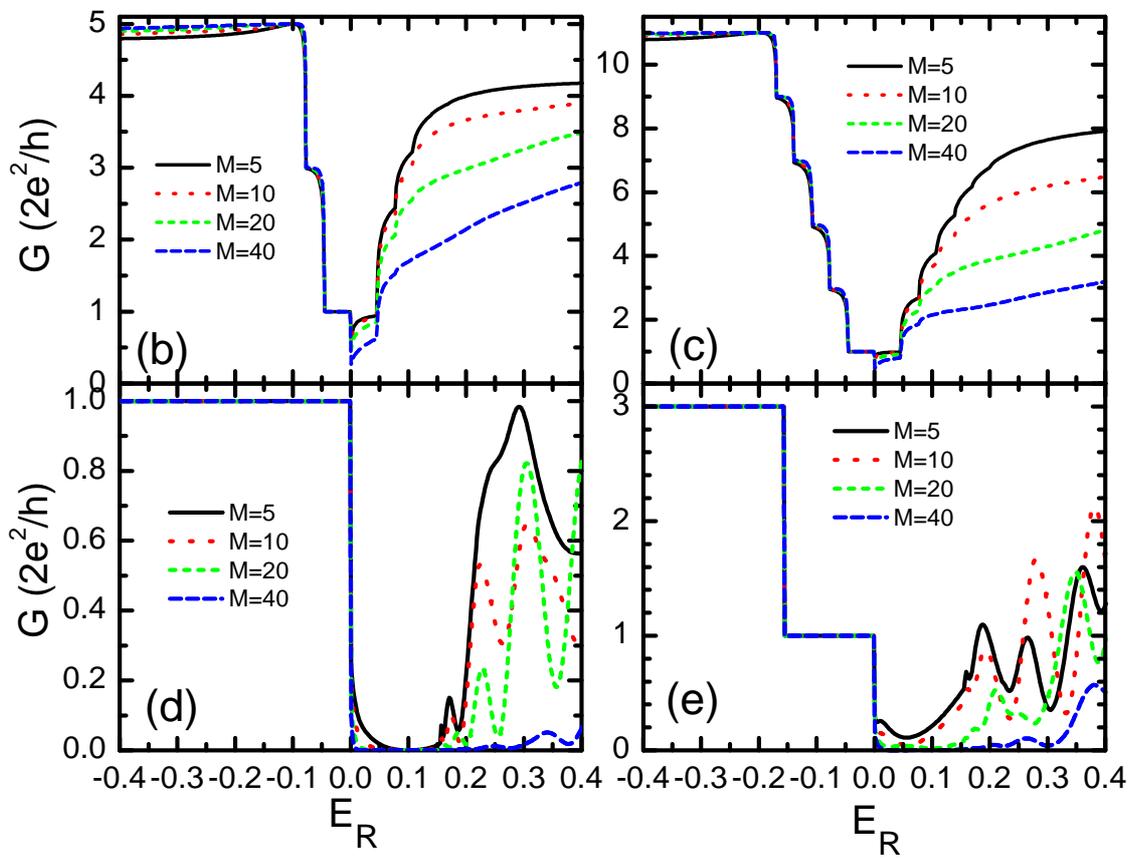
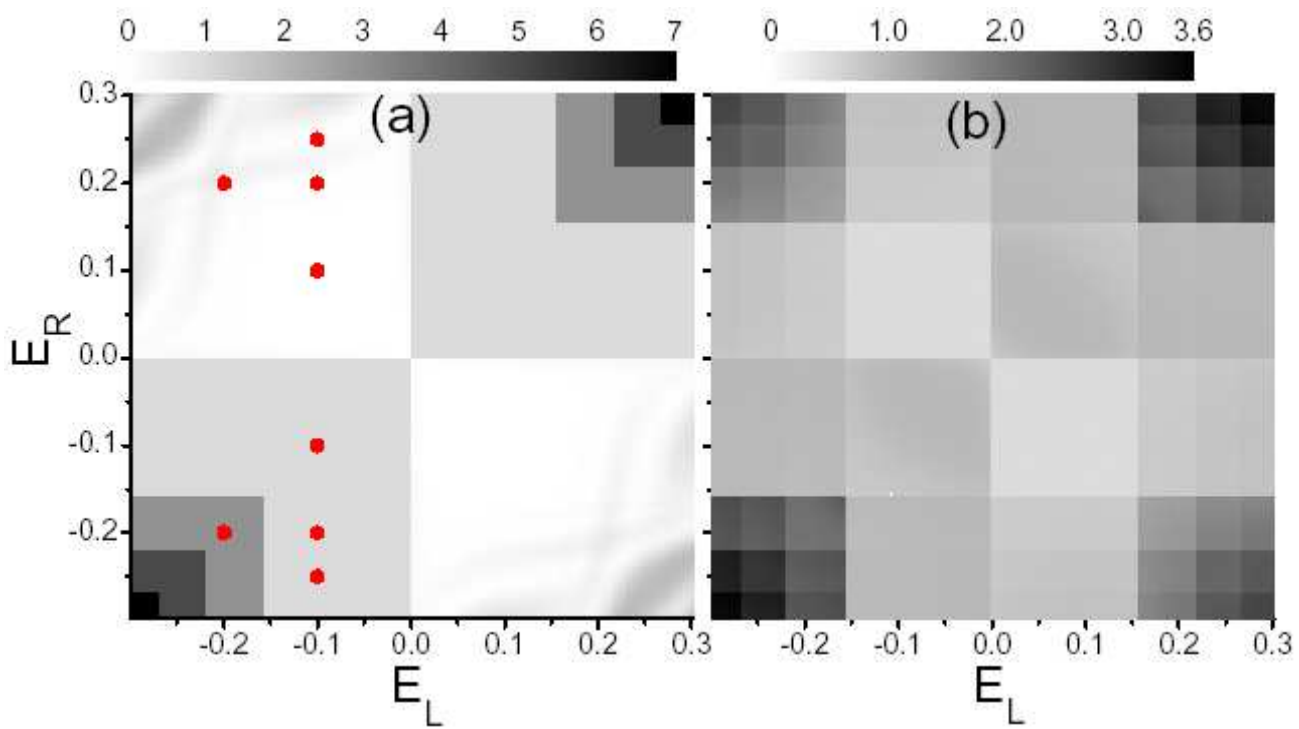
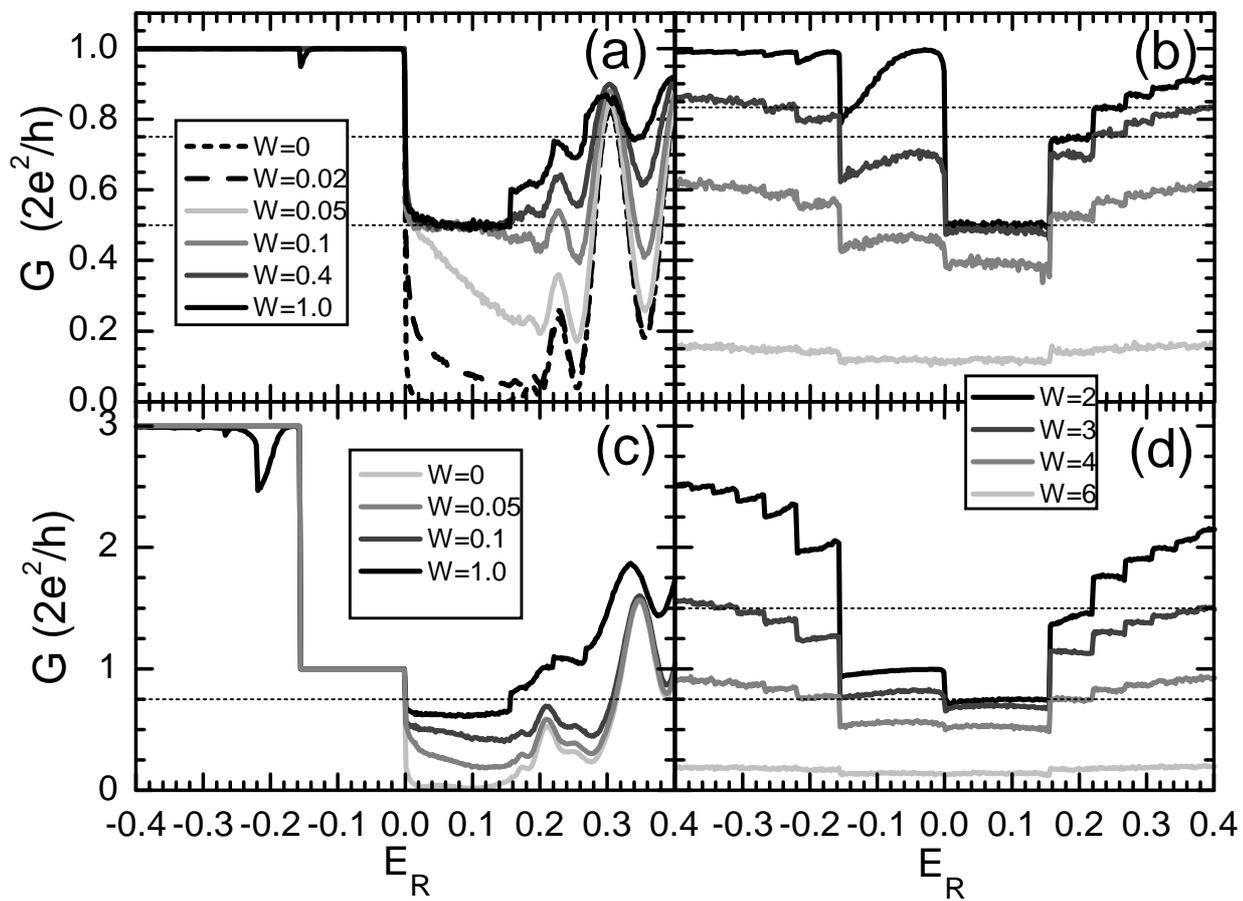
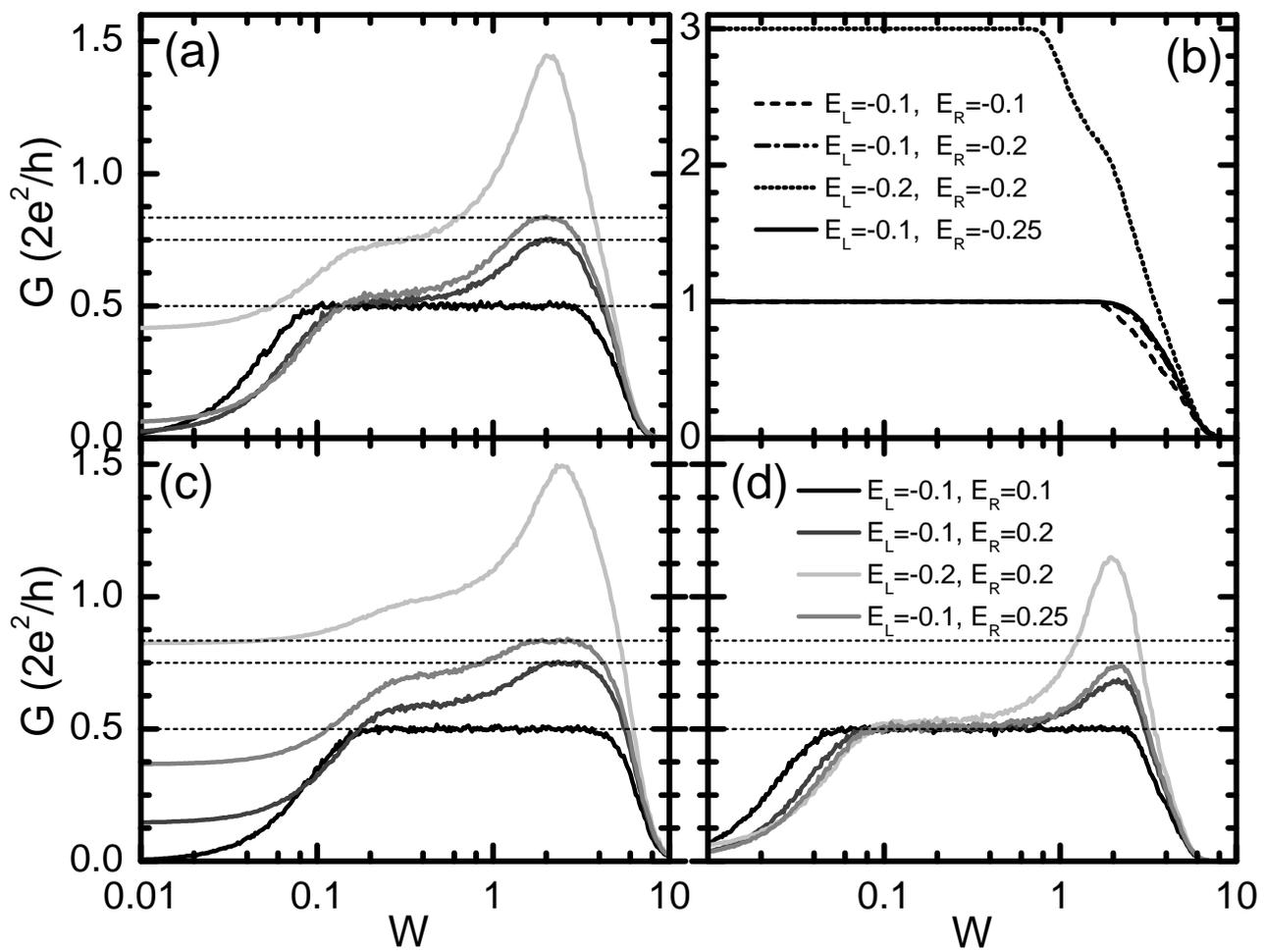
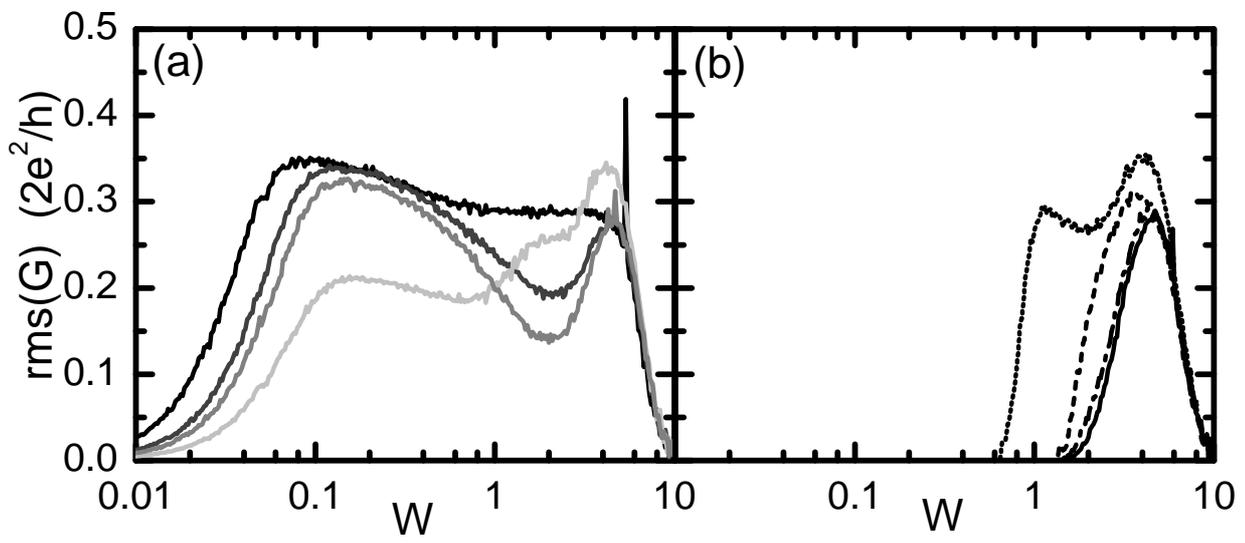
\end{document}